\documentclass[scientific reports,twocolumn,superscriptaddress]{revtex4}

\usepackage{graphicx}
\usepackage{dcolumn}
\usepackage{bm}
\usepackage{amssymb}
\usepackage{amsfonts}
\usepackage{latexsym}
\usepackage{enumerate}
\usepackage[dvipdfmx]{color} 
\usepackage{amsmath}
\usepackage[dvipdfmx]{epsfig}
\usepackage{times}
\usepackage{algorithm}
\usepackage{algorithmic}
\usepackage{cases}

\newcommand{\red}{\color{black}}
\newtheorem{theorem}{Theorem}

\newtheorem{corollary}{Corollary}
\newtheorem{definition}{Definition}

\begin{document}
\widetext
\begin{flushright}
YITP-18-103
\end{flushright}
\title{Quantum computational universality of hypergraph states with Pauli-X and Z basis measurements}
\author{Yuki Takeuchi}
\email{yuki.takeuchi.yt@hco.ntt.co.jp}
\affiliation{NTT Communication Science Laboratories, NTT Corporation, 3-1 Morinosato Wakamiya, Atsugi, Kanagawa 243-0198, Japan}
\author{Tomoyuki Morimae}
\email{tomoyuki.morimae@yukawa.kyoto-u.ac.jp}
\affiliation{Yukawa Institute for Theoretical Physics, Kyoto University, Kitashirakawa Oiwakecho, Sakyo-ku, Kyoto 606-8502, Japan}
\affiliation{JST, PRESTO, 4-1-8 Honcho, Kawaguchi, Saitama 332-0012, Japan}
\author{Masahito Hayashi}
\email{masahito@math.nagoya-u.ac.jp}
\affiliation{Graduate School of Mathematics, Nagoya University, Nagoya 464-8602, Japan}
\affiliation{Shenzhen Institute for Quantum Science and Engineering, Southern University of Science and Technology, Shenzhen 518055, China}
\affiliation{Centre for Quantum Technologies, National University of Singapore, 3 Science Drive 2 117542, Singapore}

\begin{abstract}
\noindent Measurement-based quantum computing is one of the most promising quantum computing models. Although various universal resource states have been proposed
so far,
it was open whether only two Pauli bases are enough
for both of universal measurement-based quantum computing and its verification.
In this paper, we construct a universal hypergraph state that only 
requires $X$ and $Z$-basis measurements for universal measurement-based quantum computing. 
We also show that universal measurement-based quantum computing on our hypergraph state can be verified in polynomial time using only $X$ and $Z$-basis 
measurements. Furthermore, in order to demonstrate an advantage of 
our hypergraph state, we construct a verifiable blind quantum computing 
protocol that requires only $X$ and $Z$-basis measurements for the client.
\end{abstract}
\maketitle

\noindent Quantum computing is believed to solve several 
problems faster than classical computing~\cite{S97,AAEL07,BV97}. Toward 
realizations of universal quantum computers, several quantum computing models have been proposed, such as the quantum circuit model~\cite{NC00}, adiabatic quantum computing~\cite{FGGS00}, measurement-based quantum computing (MBQC)~\cite{RB01,RBB03}, and topological quantum computing~\cite{K03}. Among these, MBQC is one of the most promising models. In MBQC, quantum computing proceeds via adaptive single-qubit measurements on a highly entangled state, a so-called universal resource state. This important advantage of MBQC, namely,
the fact that
all multi-qubit operations can be done offline, 
fits MBQC to several physical systems such as photons~\cite{SSBR15, WRRSWVAZ05}, cold atoms~\cite{BBDRN09}, ion traps~\cite{LJZHMDBBR13}, and superconducting circuits~\cite{ABSRLRS18}.

Finding fewer and simpler measurement bases
is essential for realizations of MBQC, because 
measurements are only online operations. 
Furthermore, when MBQC is applied to cloud (blind) quantum computing~\cite{MF13}, fewer and simpler measurement bases are more desirable for
the client, who securely delegates his/her quantum computing to a remote quantum server.

In this paper, we propose a universal hypergraph state that only needs Pauli-$X$ and $Z$ basis measurements (for the definition of hypergraph states, see the ``Hypergraph states" subsection in the Results section). Several previous universal resource states are summarized in Table~\ref{comparison}. We can see that except for the M\o lmer-S\o rensen graph state~\cite{KW17}, all previous universal resource states in Table~\ref{comparison} need at least three measurement bases. Our resource state is better than these resource states in the sense that our resource state only needs two Pauli-measurement bases. Although the M\o lmer-S\o rensen graph state also achieves two Pauli-measurement bases, i.e.
Pauli $X$ and $Z$ bases, an important advantage of our universal resource state is that it is efficiently verifiable with only Pauli-$X$ and $Z$ basis measurements: we can check whether a given state is close to the ideal resource state or not by measuring each qubit in the $X$ or $Z$ basis (for details, see the ``Verification of our universal hypergraph state" subsection in the Results section).
The M\o lmer-S\o rensen graph state is, on the other hand, not known to be efficiently verifiable {\red using only Pauli-$X$ and $Z$ basis measurements}.
{\red More precisely, if other additional bases than the $X$ and $Z$ bases are allowed, weighted graph states are efficiently verifiable~\cite{HT19}. These additional bases should be necessary.} This is because the stabilizer operators of weighted graph states are not linear combinations of tensor products of $X$ and $Z$ with real coefficients, and checking stabilizers seems to be the only way of efficiently verifying weighted graph states.

\begin{table*}[t]
\begin{tabular}{c|c|c}
Resource state & Measurement basis & Class \\ \hline\hline
Cluster state~\cite{BR01} & $X,Y,TXT^\dag$~\cite{MDF17} & graph state \\
Brickwork state~\cite{BFK09} & $X,Y,TXT^\dag$~\cite{BFK09} & graph state\\
Triangular lattice state~\cite{MP13} & $X,Z,H,XHX$~\cite{MP13} & graph state\\
Raussendorf-Harrington-Goyal (RHG) lattice~\cite{RHG06,RHG07} & $X,Y,Z,TXT^\dag$~\cite{RH07,RHG07} & graph state \\
Decorated RHG lattice~\cite{MF12} & $X,Y,TXT^\dag$~\cite{MF12} & graph state\\
Affleck-Kennedy-Lieb-Tasaki (AKLT) state~\cite{AKLT88,BM08} & qutrit bases~\cite{BM08} & matrix-product state\\
Union Jack state~\cite{MM16} & $X,Y,Z$~\cite{MM16} & hypergraph state\\
Three-uniform hypergraph state~\cite{GGM18} & $X,Y,Z$~\cite{GGM18} & hypergraph state\\
M\o lmer-S\o rensen graph state~\cite{KW17} & $X,Z$~\cite{KW17} & weighted graph state\\ \hline
Our state & $X,Z$ & hypergraph state
\end{tabular}
\caption{Required measurement bases for various universal resource states of MBQC. Here, $H\equiv|+\rangle\langle 0|+|-\rangle\langle 1|$ and $T\equiv|0\rangle\langle 0|+e^{i\pi/4}|1\rangle\langle 1|$, where $|\pm\rangle\equiv(|0\rangle\pm|1\rangle)/\sqrt{2}$. Operators $X$, $Y$, and $Z$ are Pauli-$X$, $Y$, and $Z$ operators, respectively. Hypergraph states are generalizations of graph states (for details, see the ``Hypergraph states" subsection in the Results section). Weighted graph states are other generalizations of graph states~\cite{HCDB07}.}
\label{comparison}
\end{table*}

In summary, our universal resource state achieves both the universality and the verifiability at the same time with only $X$ and $Z$-basis measurements.
This result should also be contrasted with the graph state case: the verification of graph states can be done with only $X$ and $Z$-basis measurements~\cite{HM15,MK18,TMMMF18}, but when we do MBQC on a graph state, an extra non-Clifford basis measurement is necessary due to the Gottesman-Knill theorem.

Our universal hypergraph state, which we denote $|G_n^d\rangle$, can simulate any $n$-qubit quantum circuit of depth $d$ consisting of the Hadamard gate $H\equiv|+\rangle\langle 0|+|-\rangle\langle 1|$ and the controlled-controlled-$Z$ $(CCZ)$ gate $CCZ\equiv I^{\otimes 3}-2|111\rangle\langle 111|$ with only adaptive $X$ and $Z$-basis measurements, where $|\pm\rangle\equiv(|0\rangle\pm|1\rangle)/\sqrt{2}$, and $I$ is the two-dimensional identity gate. Since the gate set $\{H,CCZ\}$ is universal~\cite{S02}, our state $|G_n^d\rangle$ is a universal resource state.
Our universal resource state $|G_n^d\rangle$ is constructed in the following
three steps. First, we define a hypergraph state $|G_3^1\rangle$ that can simulate any three-qubit quantum circuit of depth one consisting of $H$ and $CCZ$. Second, by entangling the $m$ hypergraph states $|G_3^1\rangle^{\otimes m}$, we construct another hypergraph state $|G_n^1\rangle$ that can simulate one-depth quantum computing on $n$ input qubits, where $m={\rm poly}(n)$. Finally, by entangling the $d$ hypergraph states $|G_n^1\rangle^{\otimes d}$, we construct the hypergraph state $|G_n^d\rangle$. In the proof of universality, we only use basic techniques of MBQC:
an $X$-basis measurement teleports a qubit 
(up to an Hadamard gate), 
and a $Z$-basis measurement decouples a qubit.

Note that in this paper we require that a resource state is a fixed state 
independent of the quantum circuit that we want to implement, because otherwise we cannot enjoy the advantage of MBQC.
For example, universal quantum 
computing is possible with $X$ and $Z$-basis measurements on
Feynman-Kitaev history states~\cite{F86,KSV02,BL08}.
However, they cannot be generated before the quantum circuit is fixed.
(More trivially, no measurement is required to simulate a quantum circuit
$U$ if our ``resource state" is $U|0\rangle^{\otimes n}$.)

Furthermore, in order to demonstrate an advantage of our hypergraph state, we propose a verifiable blind quantum computing (VBQC) protocol in which the client only needs $X$ and $Z$-basis measurements. In VBQC, a client with computationally weak quantum devices delegates universal quantum computing to a remote quantum server in such a way that the client's privacy (input, algorithm, and output) is information-theoretically protected and at the same time the honesty of the server is verifiable.
In the original proposal of VBQC~\cite{FK17}, the client has to prepare ten kinds of single-qubit states at the server's side. Although this requirement for the client has already been reduced to the $X$ and $Z$-basis measurements in Ref.~\cite{TFMI16}, our VBQC protocol is much simpler than it
(for details, see the ``Application" subsection in the Results section).
{\red The reduction of the number of measurement bases required for MBQC and its verification would be a plausible way to ease the client's burden in the sense that it seems to be impossible to make the client of VBQC completely classical as long as we require the information-theoretical security~\cite{MK19,ACGK17,MNTT18}.}\\

\medskip
\noindent{\bf\large Results}\\
This section is organized as follows: first, as preliminaries, we review the definition of hypergraph states. second, as the main result, we construct the universal hypergraph state that enables universal quantum computing with $X$ and $Z$-basis measurements. Third, we discuss the verifiability of our hypergraph state without assuming any i.i.d. property. Finally, we propose the VBQC protocol using our hypergraph state.

\medskip
\noindent{\bf Hypergraph states.}
In this subsection, we review the definition of hypergraph states~\cite{RHBM13}.
\begin{definition}[Hypergraph states]
Let $G\equiv(V,E)$ be a hypergraph, i.e. a pair of a set $V$ of vertices and a set $E$ of hyperedges, where the number $|V|$ of vertices is $n$. A hyperedge is a set of vertices. A hypergraph state $|G\rangle$ corresponding to $G$ is defined as
\begin{eqnarray*}
|G\rangle\equiv \left(\prod_{e\in E}\widetilde{CZ}_e\right)|+\rangle^{\otimes n},
\end{eqnarray*}
where each $|+\rangle$ state is placed on each vertex,
\begin{eqnarray*}
\widetilde{CZ}_e\equiv 
\bigotimes_{i\in e}I_i-2\bigotimes_{i\in e}|1\rangle\langle 1|_i
\end{eqnarray*}
is the generalized controlled-$Z$ $(CZ)$ gate acting on vertices in the hyperedge $e$, and $I_i$ is the two-dimensional identity gate on the $i$th qubit.
\end{definition}
Note that in this paper, we only consider hypergraph states satisfying $2\le|e|\le 3$ for all $e\in E$, where $|e|$ is the number of vertices in the hyperedge $e$, because such a restriction is enough for the construction
of our hypergraph state. When $|e|=2$ and $3$, the generalized $CZ$ gate becomes the conventional $CZ$ gate $CZ\equiv|0\rangle\langle 0|\otimes I+|1\rangle\langle 1|\otimes Z$ and the $CCZ$ gate, respectively. Since the $CCZ$ gate is a non-Clifford operation, hypergraph states are out of the application range of the Gottesman-Knill theorem.

\begin{figure*}[t]
\includegraphics[width=12cm, clip]{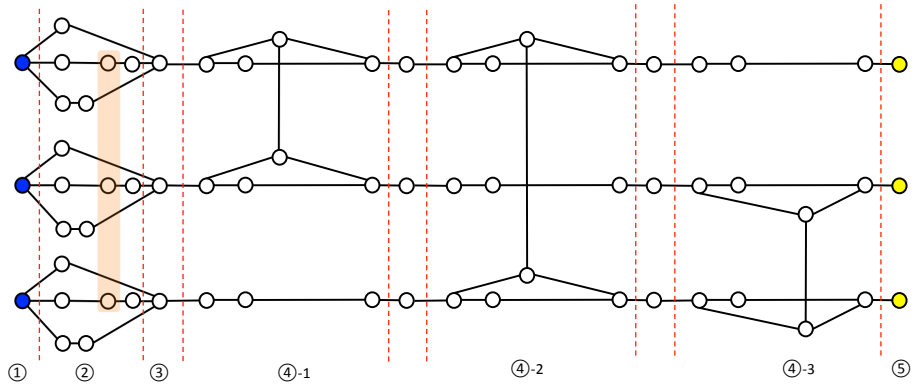}
\caption{The hypergraph $G_3^1$ to define the hypergraph state $|G_3^1\rangle$ simulating any quantum circuit of depth one consisting of $H$ and $CCZ$ on three input qubits. The orange rectangle represents a hyperedge connecting three vertices{\red, which corresponds to the $CCZ$ gate. The black lines represent edges connecting two vertices, which correspond to the $CZ$ gates.} The hypergraph $G_3^1$ has $66$ vertices and is separated into five regions. In addition, the fourth region is also separated into three parts. Each circled number represents the number of each region. Since input (blue) and output (yellow) qubits are prepared in the first and the fifth regions, we call them the input and output regions, respectively.
{\red For example, if we simulate the $CCZ$ gate using $|G_3^1\rangle$, the three input qubits corresponding to three blue vertices are finally teleported to three output qubits corresponding to three yellow vertices while $CCZ$ is applied.}}
\label{hyper}
\end{figure*}

\medskip
\noindent {\bf Pauli-X and Z universal hypergraph state.}
In this subsection, we give an intuitive idea of how to construct the hypergraph state $|G_n^d\rangle$ that enables $n$-qubit $d$-depth universal quantum computing with only adaptive $X$ and $Z$-basis measurements. First, we construct a small hypergraph state $|G_3^1\rangle$ that can simulate one-depth quantum computing on three input qubits.
Second, by entangling the $m=\binom{n}{3}$ small hypergraph states $|G_3^1\rangle^{\otimes m}$ and several single- and two-qubit states, we construct another hypergraph state $|G_n^1\rangle$ that can simulate one-depth quantum computing on $n$ input qubits. Finally, by entangling the $d$ hypergraph states $|G_n^1\rangle^{\otimes d}$ and several single-qubit states, we define the target hypergraph state $|G_n^d\rangle$ on ${\rm poly}(n,d)$ qubits that can simulate $d$-depth quantum computing on $n$ input qubits. That is, our hypergraph state $|G_n^d\rangle$ realizes universal quantum computing only with $X$ and $Z$-basis measurements.

\medskip
\noindent {\bf A hypergraph state for one-depth quantum computing on three input qubits.}
To simulate one-depth quantum computing on three input qubits, we define the hypergraph state $|G_3^1\rangle$ on $66$ qubits, based on the hypergraph $G_3^1$ defined in Fig.~\ref{hyper}.
It satisfies the following theorem:

\begin{figure}[t]
\includegraphics[width=8.5cm, clip]{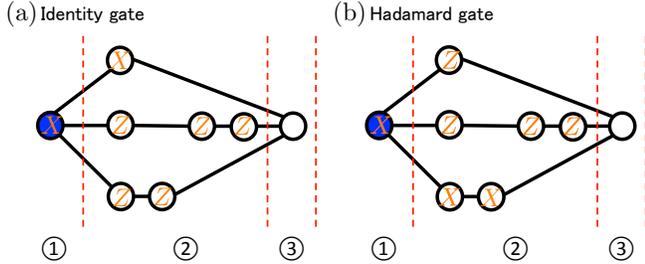}
\caption{Measurement patterns to realize single-qubit operations. This figure shows a graph state and the measurement patterns on the graph state. $X$ and $Z$ represent the $X$ and $Z$-basis measurements, respectively. (a) The measurement pattern for the identity operator. (b) The measurement pattern for the Hadamard gate.}
\label{pattern}
\end{figure}

\begin{figure}[t]
\includegraphics[width=4cm, clip]{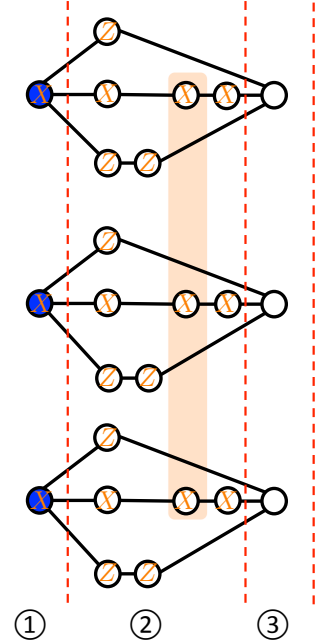}
\caption{The measurement pattern to realize the $CCZ$ gate up to nonlocal byproducts ($CZ$ gates).}
\label{pattern2}
\end{figure}

\begin{theorem}
The hypergraph state $|G_3^1\rangle$ defined by Fig.~\ref{hyper} can simulate any quantum circuit of depth one consisting of $H$ and $CCZ$ on three input qubits with adaptive $X$ and $Z$-basis measurements.
\label{universal}
\end{theorem}
{\red
{\it Proof.} Our idea is that we embed all nine patterns, $H^a\otimes H^b\otimes H^c$ and $CCZ$, where $a,b,c\in\{0,1\}$, of applying quantum gates into the hypergraph state $|G_3^1\rangle$ and then select one pattern by adaptive $X$ and $Z$-basis measurements. Below, we show Theorem~\ref{universal} according to two steps.

\medskip
{\bf Step 1:} Simulation of $H$ and $CCZ$ up to byproducts

The first region in Fig.~\ref{hyper} corresponds to three input qubits. The second region corresponds to the one-depth quantum computation. In MBQC, by measuring a qubit whose state is $|\psi\rangle$ in the $X$ basis, $|\psi\rangle$ is teleported to a neighboring qubit connected to the measured qubit while $H$ is applied on $|\psi\rangle$. In other words, the state of the neighboring qubit becomes $H|\psi\rangle$ up to a Pauli byproduct. On the other hand, by measuring a qubit in the $Z$ basis, we can decouple the qubit from other qubits. In this proof, we use these two properties of MBQC. 

First, we explain how to perform a tensor product of single qubit operations, i.e. $H_i$ $(i\in \{1,2,3\})$, $H_j\otimes H_k$ $((j,k)\in\{(1,2),(1,3),(2,3)\})$, or $H_1\otimes H_2\otimes H_3$ (Remember that we now focus on the universal gate set $\{H,CCZ\}$). These operations can be realized by the combination of the Hadamard gates and the identity operators. In Fig.~\ref{pattern}, we give the explicit measurement patterns to realize these two operations. We choose a single path by the $X$-basis measurements, and delete other two paths by the $Z$-basis measurements.
Consider for instance applying $H\otimes H\otimes I$ on input qubits $|+\rangle^{\otimes 3}$. In this case, for the first and the second input qubits, we select the lower paths. For the third input qubit, we select the upper path. As a result, from the two properties of MBQC, $(H\otimes H\otimes I)|+\rangle^{\otimes 3}$ is prepared in the third region up to byproducts. 
Since the byproducts are tensor products of $X$ and $Z$, we can remove its effect by adapting following single-qubit measurement directions.

However, when we simulate $CCZ$, byproducts include the $CZ$ gates, because $X_i(CCZ_{ijk})X_i=CCZ_{ijk}CZ_{jk}$. In order to simulate $CCZ$, we measure qubits in the first and the second regions following the measurement pattern in Fig.~\ref{pattern2}. This measurement pattern corresponds to select the middle paths for all input qubits. As a result, the state of qubits in the third region becomes $CCZ|+\rangle^{\otimes 3}$ up to byproducts including $CZ$ gates. Since $CZ$ is not a single-qubit Pauli operation, we have to correct it.

\medskip
{\bf Step 2:} Correction of nonlocal byproducts caused by applying a $CCZ$

In order to correct $CZ$ gates, we use qubits in the fourth region in Fig.~\ref{hyper}. Note that Pauli byproducts do not have to be corrected at this time because they can be accounted by adapting following single-qubit measurement directions. In the fourth region, we again use $X$ and $Z$-basis measurements to realize gate teleportations and decoupling. Now, as a byproduct, there are three kinds of $CZ$ gates, i.e. $CZ_{12}$, $CZ_{13}$, and $CZ_{23}$. The three parts of the fourth region are prepared to correct each of them. In the first part of the fourth region, there are two paths for the first and the second input qubits. If qubits in the upper paths are measured in the $X$ basis and other two qubits in the lower paths are measured in the $Z$ basis, $CZ$ is applied on the first and the second input qubits. In other words, we can correct the byproduct $CZ_{12}$. On the other hand, if we do not want to apply the $CZ_{12}$, we select the lower paths, i.e. measure qubits in lower and upper paths in the $X$ basis and the $Z$ basis, respectively. With respect to the third input qubit, by measuring qubits in the $X$ basis, the state in the third region is teleported to the region between the first and the second parts of the fourth region. The same argument holds for the second and the third parts of the fourth region to correct $CZ_{13}$ and $CZ_{23}$. Finally, the output quantum state is teleported in the fifth region up to Pauli byproducts. Therefore, we can simulate any quantum circuit of depth one consisting of $H$ and $CCZ$ on three input qubits using the hypergraph state $|G_3^1\rangle$ defined by Fig.~\ref{hyper}.
\hspace{\fill}$\blacksquare$}

\begin{figure*}[t]
\includegraphics[width=15cm, clip]{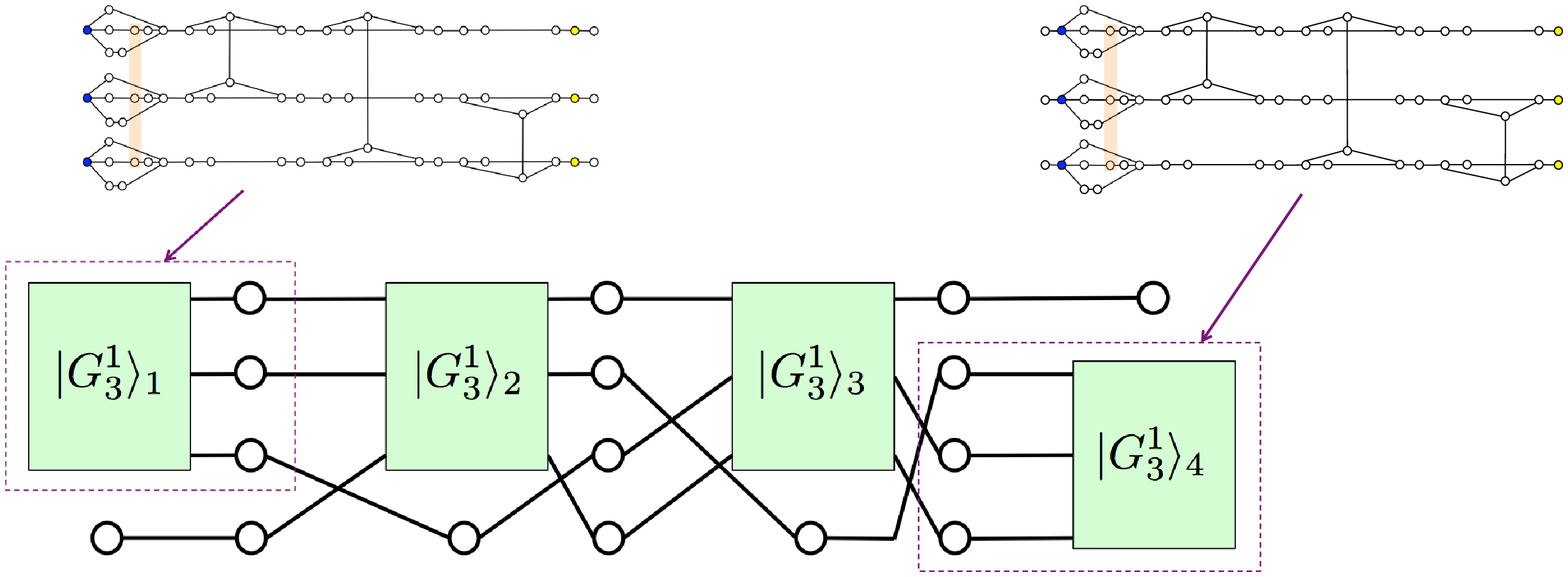}
\caption{The hypergraph $G_4^1$ to define the hypergraph state $|G_4^1\rangle$ that enables one-depth quantum computing on four input qubits using $X$ and $Z$-basis measurements. Each rectangle represents the hypergraph $G_3^1$ defined in Fig.~\ref{hyper}.}
\label{G_4}
\end{figure*}

\medskip
\noindent {\bf A hypergraph state for one-depth quantum computing on $n$ input qubits.}
In this subsection, based on the hypergraph state $|G_3^1\rangle$ given in the previous subsection, we construct another hypergraph state $|G_n^1\rangle$ that can simulate one-depth quantum computing on $n$ input qubits.
Before explaining the general construction of $|G_n^1\rangle$ for any $n$, here we explain our basic idea with a simple example of $n=4$.
When $n=4$, one-depth quantum computing means that we can apply $H_i$ $(1\le i\le 4)$, $H_i\otimes H_j$ $(1\le i<j\le 4)$, $H_i\otimes H_j\otimes H_k$ $(1\le i<j<k\le 4)$, $H_1\otimes H_2\otimes H_3\otimes H_4$, $CCZ_{ijk}$ $(1\le i<j<k\le 4)$, or $CCZ_{ijk}\otimes H_l$ $(1\le i<j<k\le 4, l\neq i,j,k)$ as our wish.
Let us consider a hypergraph state $|G_4^1\rangle$ defined by Fig.~\ref{G_4}.
From Theorem~\ref{universal}, $|G_3^1\rangle_1$, $|G_3^1\rangle_2$, $|G_3^1\rangle_3$, and $|G_3^1\rangle_4$ can be used to perform $CCZ_{123}$, $CCZ_{124}$, $CCZ_{134}$, and $CCZ_{234}$, respectively. The Hadamard gate $H$ and the identity operator $I$ can also be applied on any input qubit using the measurement pattern similar to that of Theorem~\ref{universal}. Note that the first, second, and third input qubits are included in $|G_3^1\rangle_1$. Similarly, the second, third, and fourth output qubits are included in $|G_3^1\rangle_4$.
In other words, we embed all patterns of one-depth quantum computing on four input qubits
into the hypergraph state $|G_4^1\rangle$ defined by Fig.~\ref{G_4}, and then we select a single pattern from them using $X$ and $Z$-basis measurements.
Therefore, the hypergraph state $|G_4^1\rangle$ defined by Fig.~\ref{G_4} enables one-depth quantum computing on four input qubits with $X$ and $Z$-basis measurements.

\begin{figure*}[t]
\includegraphics[width=9cm, clip]{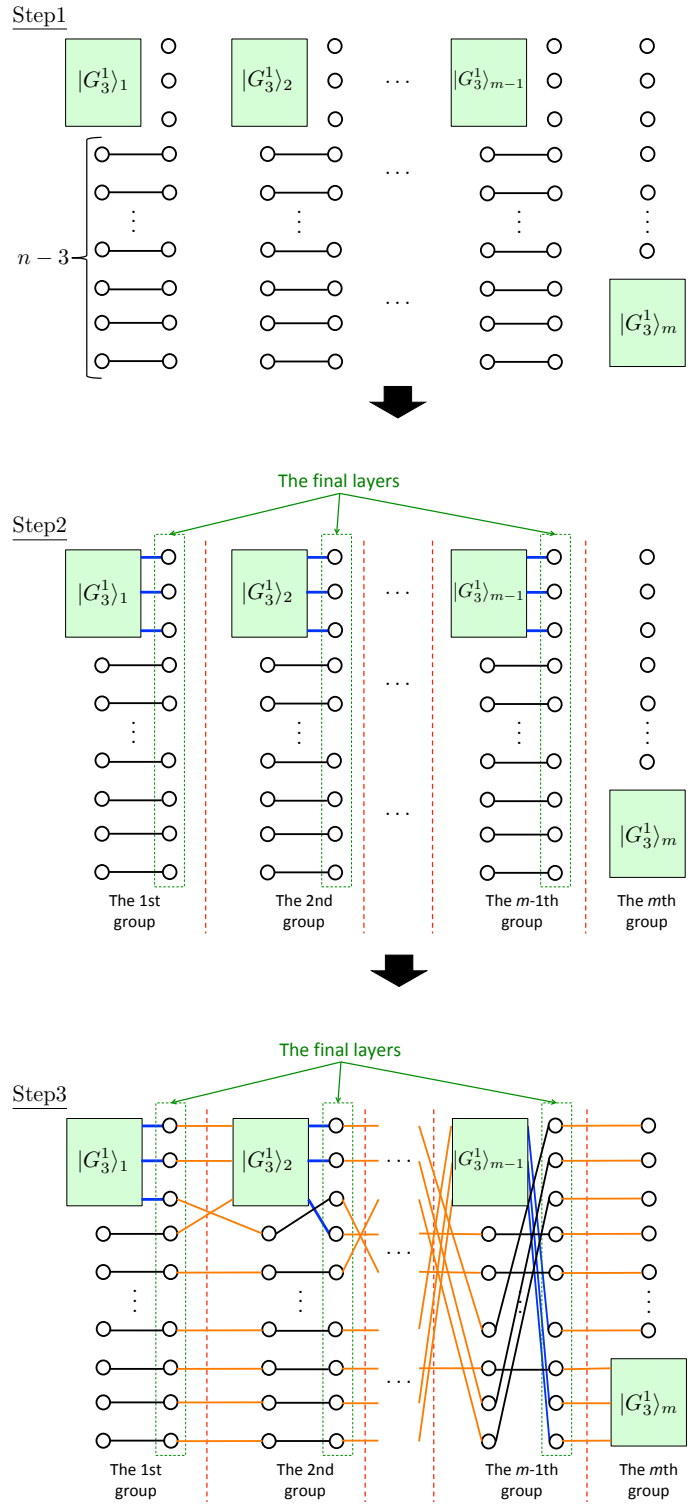}
\caption{The construction of the hypergraph $G_n^1$ to define $|G_n^1\rangle$. Each rectangle represents the hypergraph $G_3^1$ defined in Fig.~\ref{hyper}. The black, blue, and orange lines represent edges corresponding to the $CZ$ gates applied in steps 1, 2, and 3, respectively. Here, $m=\binom{n}{3}$.}
\label{construction}
\end{figure*}

For general $n$, we apply the same idea as the case of $n=4$. Using the $m=\binom{n}{3}$ hypergraph states $|G_3^1\rangle^{\otimes m}$, we define the hypergraph state $|G_n^1\rangle$ on $(2n+63)\binom{n}{3}-n$ qubits as follows:
\begin{definition}
\label{G'}
The hypergraph state $|G_n^1\rangle$ is the state constructed in the following three steps (see Fig.~\ref{construction}):
\begin{enumerate}
\item[Step 1.] Prepare $m$ states, $3(m-2)+n$ states, and $(n-3)(m-1)$ states in $|G_3^1\rangle$, $|+\rangle$, and $|\Phi^+\rangle\equiv CZ|+\rangle^{\otimes 2}$, respectively. Here, $m=\binom{n}{3}$.

\item[Step 2.] Let $|G_3^1\rangle_i$ be the $i$-th $|G_3^1\rangle$ $(1\le i\le m)$. Apply the $CZ$ gate on each qubit in the fifth region of $|G_3^1\rangle_j$ and $|+\rangle$, where $1\le j\le m-1$. For the definition of the region of $|G_3^1\rangle$, see Fig.~\ref{hyper}.

\item[Step 3.] For all $(i,j,k)$ $(1\le i<j<k\le n)$ except for $(i,j,k)=(1,2,3)$ and $(n-2,n-1,n)$, apply the $CZ$ gates on the $i$-th, $j$-th, and $k$-th qubits in the final layer of the $t$-th $(t\ge 1)$ group and the first, second, and third qubits in the first region of $|G_3^1\rangle_{t+1}$, respectively, where
\begin{eqnarray}
\nonumber
t&=&\left(\sum_{s=0}^{i-1}\sum_{l=2}^{n-1}n-l-s+1\right)+\left(\sum_{l=1}^{j-i}n-l-i+1\right)\\
\label{t}
&&+k-j-\cfrac{(n+1)(n-2)}{2}-n+i-1.
\end{eqnarray}
Note that the above operations are done such that the $w$-th $(1\le w\le n)$ qubit from the top in the final layer of the $t$-th group is connected to that in the final layer of the $(t+1)$-th group via $|G_3^1\rangle_{t+1}$.
In addition, if $t\neq m-1$, apply the $CZ$ gate on each other qubit (qubits except for the $i$-th, $j$-th, and $k$-th ones) in the final layer of the $t$-th group and a left qubit of $|\Phi^+\rangle$ in the $(t+1)$-th group. Here, the left qubit denotes a qubit that is not in the final layer.
On the other hand, if $t=m-1$, apply the $CZ$ gate on each other qubit in the final layer of the $(m-1)$-th group and $|+\rangle$ in the $m$-th group.
\end{enumerate}
\end{definition}
{\red The derivation of Eq.~(\ref{t}) is given in the Methods section.}
In Definition~\ref{G'}, when $n=4$, $(i,j,k)$ can be equal to $(1,2,4)$, $(1,3,4)$, and $(2,3,4)$. For each value, $t=1$, $2$, and $3$, respectively. Therefore, in this case, $|G_n^1\rangle$ indeed becomes the hypergraph state corresponding to $G_4^1$ shown in Fig.~\ref{G_4}.
In addition, since $|G_3^1\rangle$ is composed of $66$ qubits, from Definition~\ref{G'}, we can calculate the number of qubits in $|G_n^1\rangle$ as follows:
\begin{eqnarray}
\nonumber
&&66\binom{n}{3}+2(n-3)\left(\binom{n}{3}-1\right)+3\left(\binom{n}{3}-2\right)+n\\
\label{resource}
&=&(2n+63)\binom{n}{3}-n.
\end{eqnarray}

The hypergraph state $|G_n^1\rangle$ satisfies the following corollary:
\begin{corollary}
\label{universal2}
The hypergraph state $|G_n^1\rangle$ in Definition~\ref{G'} can simulate any quantum circuit of depth one consisting of $H$ and $CCZ$ on $n$ input qubits with adaptive $X$ and $Z$-basis measurements.
\end{corollary}
{\it Proof.} In our construction, any triple of input qubits are connected via $|G_3^1\rangle$.
{\red More precisely, for any $(i,j,k)$ except for $(1,2,3)$, there exists a single $t$ ($1\le t\le m-1$) such that the $i$th, $j$th, and $k$th qubits from the top in the final layer of the $t$th group are simultaneously connected to $|G_3^1\rangle_{t+1}$.
Note that the triple $(1,2,3)$ is already simultaneously connected in $|G_3^1\rangle_1$ using the $CCZ$ gate.
Such the $t$ is uniquely identified by Eq.~(\ref{t}).
As an example, let us consider again the case of $n=4$.
The triple $(i,j,k)$ can be chosen from $(1,2,4)$, $(1,3,4)$, and $(2,3,4)$.
As shown in Fig.~\ref{G_4}, these three triples are connected to $|G_3^1\rangle_2$, $|G_3^1\rangle_3$, and $|G_3^1\rangle_4$, respectively.
Two purple dotted enclosures in Fig.~\ref{G_4} show how to connect three triples to $|G_3^1\rangle$.
From Definition~\ref{G'}, this is true for any $n\ge 3$.}
This means that $CCZ$ and $H$ can be applied any input qubit using the measurement pattern similar to that of Theorem~\ref{universal}.
Output qubits in $|G_3^1\rangle_{\red t}$ with $1\le {\red t}\le m-1$ and qubits that do not compose $|G_3^1\rangle_{\red t}$ (i.e. white vertices in Fig.~\ref{construction}) except for ones in the $m$th group are measured in the $X$ basis. By doing so, the state of output qubits in the ${\red t}$th group is teleported to input qubits in the ${\red t}+1$th group via qubits in the final layer. Here, input qubits in the ${\red t'}$th group $(1\le {\red t'}\le m)$ are composed of three input qubits in $|G_3^1\rangle_{\red t'}$ and $n-3$ qubits (white vertices in Fig.~\ref{construction}) that are not in the final layer. On the other hand, output qubits in the ${\red t'}$th group are composed of three output qubits in $|G_3^1\rangle_{\red t'}$ and $n-3$ qubits (white vertices in Fig.~\ref{construction}) that are not in the final layer. Note that the $n-3$ qubits are input qubits as well as output qubits in the ${\red t'}$th group.
Therefore, we can simulate one-depth universal quantum computing on $n$ input qubits.
\hspace{\fill}$\blacksquare$

\begin{figure*}[t]
\includegraphics[width=10cm, clip]{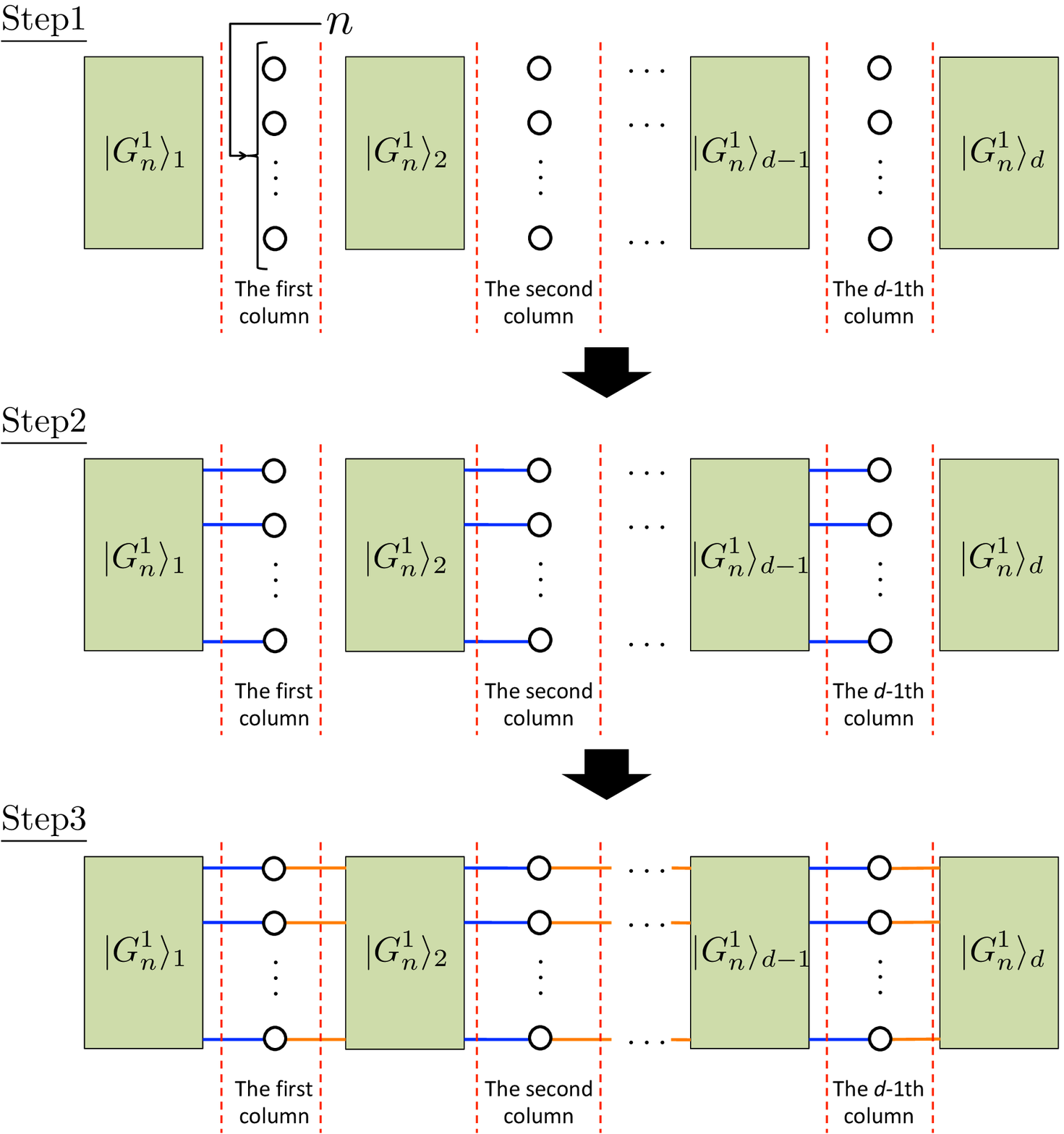}
\caption{The construction of the hypergraph $G_n^d$ to define our universal hypergraph state $|G_n^d\rangle$. Each rectangle represents the hypergraph $G_n^1$ defined in Fig.~\ref{construction}. The blue and orange lines represent edges corresponding to the $CZ$ gates applied in steps 2 and 3, respectively.}
\label{construction2}
\end{figure*}

\medskip
\noindent {\bf A hypergraph state for universal quantum computing.}
Using the hypergraph state $|G_n^1\rangle$ defined in Definition~\ref{G'}, we define the target hypergraph state $|G_n^d\rangle$ on $d(2n+63)\binom{n}{3}-n$ qubits, which enables $d$-depth universal quantum computing on $n$ input qubits with only adaptive $X$ and $Z$-basis measurements, as follows:
\begin{definition}
\label{GG}
The hypergraph state $|G_n^d\rangle$ is the state constructed in the following three steps (see Fig.~\ref{construction2}):
\begin{enumerate}
\item[Step 1.] Prepare $d$ states and $n(d-1)$ states in $|G_n^1\rangle$ and $|+\rangle$, respectively.

\item[Step 2.] Let $|G_n^1\rangle_i$ be the $i$-th $|G_n^1\rangle$. For all $t$ $(1\le t\le d-1)$, apply the CZ gate on each of $|+\rangle$ in the $t$-th column and each of the right-most qubit (output qubit) in the $m$-th group of $|G_n^1\rangle_t$ such that the $j$-th $(1\le j\le n)$ qubit from the top in the $t$-th column is connected to the $j$-th output qubit from the top in the $m$-th group of $|G_n^1\rangle_t$. For the definition of the group of $|G_n^1\rangle$, see Fig.~\ref{construction}. Here, $m=\binom{n}{3}$.

\item[Step 3.] For all $t$ $(1\le t\le d-1)$, apply the CZ gate on each of $|+\rangle$ in the $t$-th column and each of the left-most qubit (input qubit) in the first group of $|G_n^1\rangle_{t+1}$ such that the $j$-th $(1\le j\le n)$ qubit from the top in the $t$-th column is connected to the $j$-th input qubit from the top in the first group of $|G_n^1\rangle_{t+1}$.
\end{enumerate}
\end{definition}
Combining Eq.~(\ref{resource}) and Definition~\ref{GG}, we can calculate the number of qubits in $|G_n^d\rangle$ as follows:
\begin{eqnarray}
\nonumber
&&d\left[(2n+63)\binom{n}{3}-n\right]+n(d-1)\\
\label{number}
&=&d(2n+63)\binom{n}{3}-n.
\end{eqnarray}
Hence, when $d={\rm poly(n)}$, the number of vertices of $G_n^d$ is ${\rm poly}(n)$. Furthermore, the maximal number of vertices connected to a hyperedge is upper bounded by a constant number, which is three in our case. Therefore, the number of hyperedges is also ${\rm poly}(n)$.
In short, our hypergraph state $|G_n^d\rangle$ is generated in ${\rm poly}(n)$ time from ${\rm poly}(n)$ qubits when $d={\rm poly}(n)$.

Our universal hypergraph state $|G_n^d\rangle$ satisfies the following theorem:
\begin{theorem}
\label{universal3}
The hypergraph state $|G_n^d\rangle$ in Definition~\ref{GG} can simulate any quantum circuit of depth $d$ consisting of $H$ and $CCZ$ on $n$ input qubits with adaptive $X$ and $Z$-basis measurements.
\end{theorem}
{\it Proof.} Using the teleportation by the $X$-basis measurement, we can teleport the state of output qubits of $|G_n^1\rangle_t$ to input qubits of $|G_n^1\rangle_{t+1}$ via qubits in the $t$th column. Therefore, from Corollary~\ref{universal2}, $|G_n^d\rangle$ enables $d$-depth universal quantum computing on $n$ input qubits.
\hspace{\fill}$\blacksquare$

Theorem~\ref{universal3} shows that our hypergraph state $|G_n^d\rangle$ only requires $X$ and $Z$-basis measurements to realize universal quantum computing.
In other words, when $d={\rm poly}(n)$, we can efficiently simulate any quantum circuit of depth $d$ consisting of ${\rm poly}(n)$ number of $H$ and $CCZ$ by measuring our hypergraph state $|G_n^d\rangle$ in the $X$ and $Z$ bases.

\medskip
\noindent {\bf Verification of our universal hypergraph state.}
In this subsection, we discuss the verifiability of our hypergraph state $|G_n^d\rangle$ without assuming any independent and identically distributed (i.i.d.) property.
Let us consider the following general situation: there exist two parties, Alice and Bob. Alice can only perform single-qubit measurements and has no quantum memory. On the other hand, Bob possesses a universal quantum computer, i.e. he can prepare universal resource states. Therefore, she delegates the preparation of universal resource states to Bob, and he sends each qubit of his generated state one by one to Alice. However, since she does not trust him, she has to verify the correctness of his state.

\begin{figure}[t]
\includegraphics[width=7cm, clip]{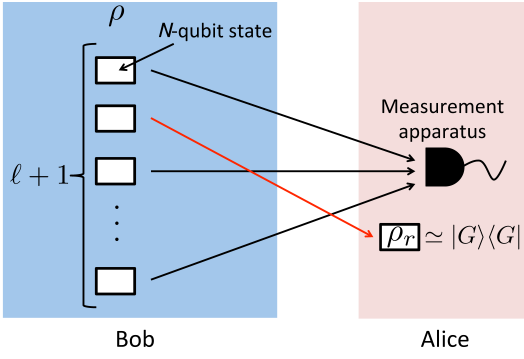}
\caption{The verification protocol for an $N$-qubit hypergraph state $|G\rangle$. First, Bob generates an $N(\ell+1)$-qubit state $\rho$, and sends each qubit of them to Alice one by one. The quantum state $\rho$ consists of $\ell+1$ registers, and each register stores $N$ qubits. Second, Alice uniformly
randomly chooses $\ell$ registers and tests them. If all the measurement outcomes satisfy Eq.~(\ref{stabilizer}), the remaining single register $\rho_r$ is guaranteed to be close to the ideal hypergraph state $|G\rangle$ (for details, see Theorem~\ref{ZH}). Therefore, Alice can safely use $\rho_r$ 
for her MBQC.}
\label{verification}
\end{figure}

Let us assume that we want to verify an $N$-qubit hypergraph state $|G\rangle$ corresponding to the hypergraph $G$.
So far, several verification protocols for hypergraph states have been proposed~\cite{MTH17,TM17,ZH18}. 
{\red All of them require only Pauli-$X$ and $Z$ basis measurements.
This is a reason why we tried to construct a universal hypergraph state that only requires Pauli-$X$ and $Z$ basis measurements for universal MBQC.}
The most resource-efficient one~\cite{ZH18} is the following protocol
(see Fig.~\ref{verification}):

\begin{enumerate}
\item Bob generates an $N(\ell+1)$-qubit state $\rho$ and sends each qubit to Alice one by one, where $N=d(2n+63)\binom{n}{3}-n$ in our case (see Eq.~(\ref{number})). If Bob is honest, $\rho=(|G\rangle\langle G|)^{\otimes \ell+1}$. If Bob is malicious, $\rho$ is an arbitrary quantum state, which may be entangled. Without loss of generality, we can assume that $\rho$ consists of $\ell+1$ registers, and each register stores $N$ qubits.
\item 
Alice uniformly randomly chooses $\ell$ registers from $\ell+1$ registers.  
For each of them, Alice applies the following protocol, the so-called cover protocol~\cite{ZH18}: Alice uniformly randomly chooses the value of $i$ from $\{1,2,\ldots,\chi(G)\}$ 
and then applies the $i$th color test. 
(Definitions of $\chi(G)$ and the $i$th color test are given later.)
If all $\ell$ registers pass the tests, Alice proceeds the next step. Otherwise, Alice aborts the protocol.
\item Alice uses the remaining single register $\rho_r$ to perform MBQC. 
\end{enumerate}

Let us define $\chi(G)$. It is the chromatic number of $G$, i.e.
the minimal number of colors in any hypergraph coloring of $G$.
Here, a hypergraph coloring is a way of coloring each
vertex such that no hyperedge contains two vertices having the same color.
We say that $G$ is $\chi(G)$-colorable. 
Let $C_i$ $(1\le i\le\chi(G))$ be the set of vertices colored by the $i$th color. 
By definition, any two elements of $C_i$ are disconnected to each other,
and $\cup_{i=1}^{\chi(G)}C_i=V$, where $V$ is the set of vertices of $G$. 

We also define the $i$th color test. 
\begin{description}
\item[$i$th color test]
We measure qubits in $C_i$ in the $X$ basis and other qubits in the $Z$ basis.
These measurements correspond to measure stabilizers of $|G\rangle$. 
Let $o_j$ be the measurement outcome on the $j$th $(1\le j\le N)$ qubit. If
\begin{eqnarray}
\label{stabilizer}
o_j+\sum_{e\in E|e\ni j}\prod_{k\in e,k\neq j}o_k\equiv0\ ({\rm mod}\ 2)
\end{eqnarray}
for all $j\in C_i$, 
we consider that the $i$th test is passed.
Here, $e\in E|e\ni j$ means the summation over hyperedges that include the $j$th vertex. Eq.~(\ref{stabilizer}) means that the tested register is properly stabilized. 
\end{description}

This verification protocol satisfies the following theorem:
\begin{theorem}[Ref.~\cite{ZH18}]
\label{ZH}
Let
\begin{eqnarray*}
\ell=\left\lceil\cfrac{\chi(G)(1-\delta)}{\delta\epsilon}\right\rceil,
\end{eqnarray*}
where $\lceil\cdot\rceil$ is the ceiling function. Then, if Alice proceeds step 3, with significance level $\delta$,
\begin{eqnarray}
\label{fidelity}
\langle G|\rho_r|G\rangle\ge 1-\epsilon.
\end{eqnarray}
\end{theorem}
Note that the significance level is the maximum probability of Alice proceeding step 3 when the state $\rho_r$ does not satisfy Eq.~(\ref{fidelity}).
This theorem means that we can estimate the lower bound of the fidelity $\langle G|\rho_r|G\rangle$ between the given state $\rho_r$ and the target hypergraph state $|G\rangle$.

\begin{figure}[t]
\includegraphics[width=8cm, clip]{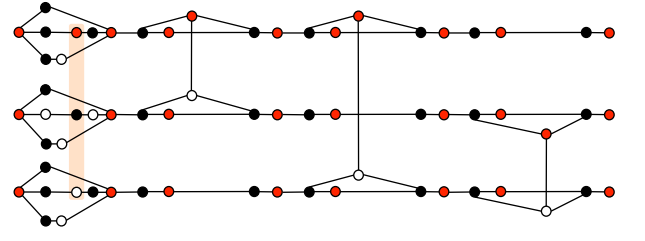}
\caption{A three-coloring of $G_3^1$. Most importantly, six vertices corresponding to the input and output qubits are colored in the same color (red). This property is useful to show $\chi(G_n^d)=3$.}
\label{coloring}
\end{figure}

Now let us apply the verification protocol to our hypergraph state $|G_n^d\rangle$.
To derive the number of registers required to verify our hypergraph state $|G_n^d\rangle$, we show the following theorem:
\begin{theorem}
\label{color}
Let $G_n^d$ be a hypergraph defined in Definition~\ref{GG}. Then,
\begin{eqnarray*}
\chi(G_n^d)=3.
\end{eqnarray*}
\end{theorem}
{\it Proof.} Since the maximum order of hyperedges of $G_n^d$ is three, $\chi(G_n^d)\ge 3$. The task left is to show that three colors are sufficient for coloring of $G_n^d$. To this end, first, we give an explicit coloring of $G_3^1$ in Fig.~\ref{coloring}. From Fig.~\ref{coloring}, it is evident that three colors (red, black, and white) are sufficient for coloring of $G_3^1$. Next, based on this coloring, we consider the coloring of $G_n^1$. In the construction of $|G_n^1\rangle$ (see Fig.~\ref{construction}), we use quantum states $|G_3^1\rangle$, $|\Phi^+\rangle$, and $|+\rangle$. Each $|G_3^1\rangle$ is colored in the same manner as in Fig.~\ref{coloring}. Since input and output qubits of $|G_3^1\rangle$ are colored in the same color (red), each single qubit $|+\rangle$ except for them in the $m$th group and each right qubit of $|\Phi^+\rangle$ can be colored in black (or white). In this case, the other qubits can be colored in red. Therefore, $G_n^1$ is also three-colorable. Furthermore, from this coloring, we notice that input and output qubits of $|G_n^1\rangle$ are also colored in the same color (red). As a result, in the construction of $|G_n^d\rangle$ (see Fig.~\ref{construction2}), all single qubits in columns can be colored in black (or white). This means that $\chi(G_n^d)=3$.
\hspace{\fill}$\blacksquare$

From Theorem~\ref{ZH} and Theorem~\ref{color}, we obtain
\begin{corollary}
Our hypergraph state $|G_n^d\rangle$ can be verified 
using $\lceil3(1-\delta)/(\delta\epsilon)\rceil+1$ samples (registers). 
\label{morimae}
\end{corollary}
It is important
to point out that the chromatic number of the Union Jack state is also 
three~\cite{ZH18}, which means that the efficiency of verifying
our hypergraph state
is the same as that of verifying the Union Jack state.

In short, from Theorem~\ref{universal3},
Corollary~\ref{morimae},
and the fact that the existing polynomial-time verification protocols~\cite{MTH17,TM17,ZH18} only require $X$ and $Z$-basis measurements, the following corollary holds:
\begin{corollary}
\label{XZall}
Our hypergraph state $|G_n^d\rangle$ is a polynomial-time generated quantum state such that $X$ and $Z$-basis measurements are sufficient 
\begin{enumerate}
\item to simulate universal quantum computing in polynomial time of $n$, and
\item to estimate the fidelity in polynomial time of the size of the hypergraph state.
\end{enumerate}
\end{corollary}
Importantly, all existing other universal resource states require at least 
one additional measurement basis other than $X$ and $Z$ bases, while
Corollary~\ref{XZall} says that our hypergraph state
only needs $X$ and $Z$-basis 
measurements
for both MBQC and verification. It is an important advantage
of our hypergraph state.

\medskip
\noindent {\bf Application.}
In this subsection, utilizing our hypergraph state $|G_n^d\rangle$, we propose a VBQC protocol where requirements for a client is only $X$ and $Z$-basis measurements.
In VBQC, a client with computationally weak quantum devices delegates universal quantum computing to a remote quantum server in such a way that the client's privacy (input, algorithm, and output) is information-theoretically protected and at the same time the honesty of the server is verifiable.

Our VBQC protocol runs as follows (the protocol is described for an honest server. If the server is malicious, the server can do any deviation that does not violate the no-signaling principle.):
\begin{enumerate}
\item The client chooses the value of $\ell'$ from $\{1,2,\ldots,\ell+1\}$ uniformly at random.
\item The quantum server generates $\ell+1$ hypergraph states $|G_n^d\rangle^{\otimes \ell+1}$, and sends each qubit of them to the client one by one.
\item For the $\ell'$th hypergraph state, the client performs MBQC. For other $\ell$ hypergraph states, the client applies the cover protocol of the ``Verification of our universal hypergraph state" subsection.
\item If the $\ell$ hypergraph states pass the cover protocol, the client accepts the outcome of the MBQC on the $\ell'$th hypergraph state. Otherwise, the client rejects the outcome.
\end{enumerate}
From the universality of our hypergraph state $|G_n^d\rangle$ (Theorem~\ref{universal3}), the client can obtain the correct result if the server is honest, i.e. the sever sends the ideal states.
Since there exists only one-way communication from the server to the client in our VBQC protocol, the privacy of the client is information-theoretically preserved due to the no-signaling principal.
The verifiability of our VBQC protocol is automatically satisfied by the verification protocol.

In the original proposal of VBQC~\cite{FK17}, the client has to prepare ten kinds of single-qubit states at the server's side.
Although this requirement for the client has already been reduced to the $X$ and $Z$-basis measurements in Ref.~\cite{TFMI16}, our VBQC protocol is simpler than it, and our constructive method is completely different from theirs. 
In particular, the security proof of our VBQC protocol is much simpler because our VBQC protocol needs only one-way communication and therefore the security is trivially satisfied due to the no-signaling principle, while the protocol of Ref.~\cite{TFMI16} needs two-way communications.
{\red Furthermore, the verifiability of the protocol in Ref.~\cite{TFMI16} has been shown by appropriately tailoring the verifiability proof of the original proposal~\cite{FK17} to this protocol.
On the other hand, the verifiability of our VBQC protocol is automatically shown from the verifiability of hypergraph states.
This difference also makes our VBQC protocol simpler than that in Ref.~\cite{TFMI16}.}

{\red Note that there are mainly two types of VBQC protocols.
In the first type, the client prepares single-qubit states and then sends them to the server. All of measurements are delegated to the server.
This type includes the original VBQC protocol~\cite{FK17}. 
On the other hand, in the second type~\cite{MF13}, the preparation of universal resource states is delegated to the server.
The only requirements for the client are single-qubit measurements on the received resource states.
Our VBQC protocol is constructed by applying our hypergraph state to the second-type VBQC protocol.}\\

\medskip
\noindent{\bf\large Discussion}\\
We have constructed a hypergraph state that enables universal quantum computing with only $X$ and $Z$-basis measurements. We have also shown that these {\red two} measurements are sufficient to verify our hypergraph state, and have also constructed a VBQC protocol, which only requires $X$ and $Z$-basis measurements for the client. Our result decreases the number of measurement bases required for reliable MBQC from existing universal resource states.
{\red Two seems to be the optimal solution, because no feed-forward operation is possible if only a single measurement basis is used.}

In this paper, we have considered a hypergraph state. It is open whether two measurement bases are enough also for a (an unweighted) graph state.
It might be possible to find a measurement pattern on a graph state where only two measurement bases are enough for universal MBQC. However, we point out that if we require that the MBQC is verifiable at the same time and use one of existing verification protocols~\cite{HM15,MK18,TMMMF18}, at least three measurement bases should be necessary. The reason is as follows. As far as we know, all of existing verification protocols require $X$ and $Z$-basis measurements for the verification of graph states. However, due to the Gottesman-Knill theorem, only $X$ and $Z$-basis measurements are not enough for the universality.
Therefore, at least a single non-Clifford basis measurement should be added, and in total, three measurement bases are necessary.
Our resource state realizes universal MBQC and its verification with the minimum number of local measurements when we restrict verification protocols to be one of existing protocols.
(Also remember that in the case of weighted graph states, $X$ and $Z$-basis measurements are enough for the universality~\cite{KW17}, but they do not seem to be enough for the verification.)

Our hypergraph state has two advantages over the Union Jack state, which is another universal hypergraph state~\cite{MM16}.
First, MBQC on our hypergraph state uses no imaginary number, while that on the Union Jack state does, because it needs $Y$-basis measurements. This feature should simplify the further theoretical analysis of MBQC on hypergraph states.
{\red As a concrete example, let us consider the self-testing. Simply speaking, the self-testing is a device-independent verification for a set of a state and operators. To devise a self-testing protocol, a certain equivalence between two sets of a state and operators is defined.
A definition of equivalence used, for example, in Ref.~\cite{M16} is tailored for the case where the ideal set of a state and operators contains no imaginary number.
Although the definition can be extended to the imaginary-number case~\cite{MM10}, the MBQC on our universal hypergraph state does not need such the extension.
This property may facilitate the proposal of a self-testing protocol for a set of our hypergraph state and Pauli-$X$ and $Z$ basis measurements.}
Second, our proof of universality is simpler than their proof, because they use the percolation argument, while we only use the basic techniques of MBQC (i.e. the teleportation by the $X$-basis measurement and the decoupling by the $Z$-basis measurement).
{\red Note that the deterministic universality (i.e., universality without the percolation argument) has already been achieved in Ref.~\cite{GGM18}. As an advantage of their universal hypergraph state over ours, their state enables to parallelize the $CCZ$ and SWAP gates while it requires more number of measurement bases than ours (i.e., three Pauli bases). It would be an interesting future work to consider whether we can construct a resource state that has both of our and their advantages without losing the deterministic universality.}

{\red At the last of this section, we mention the relation between the complexity of resource states and the required number of measurement bases for the universality.
The cluster state~\cite{BR01} only has the $CZ$ gates and requires two Pauli-basis measurements and one additional non-Clifford measurement for the deterministic universality~\cite{MDF17} (see Table~\ref{comparison}).
The Union Jack state only has the $CCZ$ gates and requires three Pauli-basis measurements for the probabilistic universality due to the percolation argument~\cite{MM16}.
This probabilistic universality can be improved to the deterministic one using another hypergraph state in Ref.~\cite{GGM18} that also only has the $CCZ$ gates and requires only three Pauli-basis measurements for the univeisality.
Our hypergraph states uses the $CZ$ and $CCZ$ gates, but the universality (and the verifiability) is achieved using only Pauli-$X$ and $Z$ basis measurements.
The M\o lmer-S\o rensen graph state, which is a weighted graph state, only uses one type of two-qubit phase gates and requires only Pauli-$X$ and $Z$ basis measurements for the deterministic universality~\cite{KW17}.
Furthermore, the cluster state, the Union Jack state, and the M\o lmer-S\o rensen graph state are embedded in two-dimensional space while ours and the hypergraph state in Ref.~\cite{GGM18} are not.
Roughly speaking, these known facts including our present result indicate that making a resource state somewhat complex gives an advantage in the number of measurement bases.
However, the relation between the complexity of resource states and the number of measurement bases seems to, in general, depend on the construction of universal resource states.
Therefore, although the complete characterization of the relation would be an interesting direction of research, it seems to be hard, which is beyond the scope of our current research.}
\\

\medskip
\noindent{\bf\large Methods}\\
{\red In this section, we derive Eq.~(\ref{t}) in Definition~\ref{G'}.
We would like to define $t$ such that it increases one by one from $0$ as $(i,j,k)$ becomes $(1,2,3)$,$(1,2,4)$,$(1,2,5)$,$\ldots$,$(1,2,n)$,$(1,3,4)$,$\ldots$,$(1,3,n)$, $\ldots$,$(1,n-1,n)$,$(2,3,4)$,$\ldots$,$(n-2,n-1,n)$. In other words, the number $t$ specifies the distance between $(i,j,k)$ and $(1,2,3)$. For example, since $(1,2,4)$ is the next triple of $(1,2,3)$, the number $t$ must be one when $(i,j,k)=(1,2,4)$. In order to ease the derivation, here we define $\tilde{t}\equiv t+1$.

First, we consider the case of $(i,j,k)=(1,2,k)$ $(3\le k\le n)$. In this case, $\tilde{t}=k-2$. In the case of $(i,j,k)=(1,3,k)$ $(4\le k\le n)$, $\tilde{t}=(n-2)+(k-3)$. Here, $(n-2)$ is the value of $\tilde{t}$ of the triple $(1,2,n)$. In the same way, when $(i,j,k)=(1,4,k)$ $(5\le k\le n)$, $\tilde{t}=(n-2)+(n-3)+(k-4)$.

Second, repeating the same calculation, when $(i,j,k)=(1,j,k)$ and $3\le j<k\le n$,
\begin{eqnarray}
\label{M1}
t=\tilde{t}-1=\left[\sum_{l=2}^{j-1}(n-l)\right]+(k-j)-1.
\end{eqnarray}
In order to take the case of $j=2$ into consideration, we slightly modify Eq.~(\ref{M1}) as follows:
\begin{eqnarray}
\nonumber
t&=&\left[\sum_{l=2}^{j-1}(n-l)\right]+(k-j)-1\\
\nonumber
&=&\left[\sum_{l=1}^{j-1}(n-l)\right]+(k-j)-1-(n-1)\\
\label{M2}
&=&\left[\sum_{l=1}^{j-1}(n-l)\right]+k-j-n.
\end{eqnarray}
Indeed, if we substitute $i=1$ for Eq.~(\ref{t}), Eq.~(\ref{M2}) is derived.

Finally, we derive $t$ for any $(i,j,k)$ with $1\le i<j<k\le n$. To this end, we use the similar argument as above. When $(i,j,k)=(2,3,k)$ $(4\le k\le n)$,
\begin{eqnarray*}
t=\left\{\left[\sum_{l=1}^{n-2}(n-l)\right]+n-(n-1)-n\right\}+(k-3),
\end{eqnarray*}
where the first term is the value of $t$ of the triple $(1,n-1,n)$.
Then, when $(i,j,k)=(2,4,k)$ $(5\le k\le n)$,
\begin{eqnarray*}
t=\left\{\left[\sum_{l=1}^{n-2}(n-l)\right]-(n-1)\right\}+(n-3)+(k-4).
\end{eqnarray*}
Therefore, using the similar argument as the case of $(i,j,k)=(1,j,k)$, when $(i,j,k)=(2,j,k)$ with $3\le j<k\le n$,
\begin{eqnarray*}
t=&&\left\{\left[\sum_{l=1}^{n-2}(n-l)\right]-(n-1)\right\}\\
&+&\left\{\left[\sum_{l=1}^{j-2}(n-l-1)\right]+(k-j)-(n-2)\right\}.
\end{eqnarray*}
Similarly, when $(i,j,k)=(3,j,k)$ with $4\le j<k\le n$,
\begin{eqnarray*}
t=&&\left\{\left[\sum_{l=1}^{n-2}(n-l)\right]-(n-1)\right\}\\
&+&\left\{\left[\sum_{l=1}^{n-3}(n-l-1)\right]-(n-3)\right\}\\
&+&\left\{\left[\sum_{l=1}^{j-3}(n-l-2)\right]+(k-j)-(n-3)\right\}.
\end{eqnarray*}
Repeating this calculation for $i$, we can obtain Eq.~(\ref{t}).}

\medskip
\noindent{\bf\large Acknowledgements}\\
We thank Jingfang Zhou and Atul Mantri for helpful discussions.
We also thank Mariami Gachechiladze, Otfried G\"{u}hne, and Akimasa Miyake for pointing out Ref.~\cite{KW17} to us.
T. M. is supported by the JSPS Grant-in-Aid for Young Scientists (B) No.JP17K12637, and JST PRESTO No.JPMJPR176A.
M. H. is supported in part by Fund for the Promotion of Joint International Research (Fostering Joint International Research) Grant No. 15KK0007, Japan Society for the Promotion of Science (JSPS) Grant-in-Aid for Scientific Research (A) No. 17H01280, (B) No. 16KT0017, and Kayamori Foundation of Informational Science Advancement.\\

\medskip
\noindent{\bf\large Author contributions}\\
Y.T. conceived the key idea. Y.T. refined the key idea through discussions with T.M. and M.H. Y.T. performed the calculations, and all the authors contributed to checking the validity of the calculations and writing of the paper.\\

\medskip
\noindent{\bf Competing interests:}
The authors declare no competing interests.

\medskip
\noindent{\red{\bf Data availability:}
No data sets were generated or analyzed during the current study.}

\begin{thebibliography}{99}
\bibitem{S97}Shor, P. W. Polynomial-Time Algorithms for Prime Factorization and Discrete Logarithms on a Quantum Computer. {\it SIAM J. Comput.} {\bf 26}, 1484 (1997).
\bibitem{AAEL07}Aharonov, D., Arad I., Eban, E. \& Landau, Z. Polynomial Quantum Algorithms for Additive approximations of the Potts model and other Points of the Tutte Plane, Preprint at arXiv:quant-ph/0702008 (2007).
\bibitem{BV97}Bernstein, E. \& Vazirani, U. Quantum complexity theory, {\it SIAM J. Comput.} {\bf 26}, 1411 (1997).
\bibitem{NC00}Nielsen, M. A. \& Chuang, I. L. {\it Quantum Computation and Quantum Information} (Cambridge University Press, Cambridge, 2000).
\bibitem{FGGS00}Farhi, E., Goldstone, J., Gutmann, S. \& Sipser, M. Quantum Computation by Adiabatic Evolution, Preprint at arXiv:quant-ph/0001106 (2000).
\bibitem{RB01}Raussendorf, R. \& Briegel, H. J. A One-Way Quantum Computer, {\it Phys. Rev. Lett.} {\bf 86}, 5188 (2001).
\bibitem{RBB03}Raussendorf, R., Browne, D. E. \& Briegel, H. J. Measurement-based quantum computation on cluster states, {\it Phys. Rev. A} {\bf 68}, 022312 (2003).
\bibitem{K03}Kitaev, A. Fault-tolerant quantum computation by anyons, {\it Ann. Phys.} {\bf 303}, 2 (2003).
\bibitem{SSBR15}G.-Segovia, M., Shadbolt, P., Browne, D. E. \& Rudolph, T. From Three-Photon Greenberger-Horne-Zeilinger States to Ballistic Universal Quantum Computation, {\it Phys. Rev. Lett.} {\bf 115}, 020502 (2015).
\bibitem{WRRSWVAZ05}Walther, P. {\red{\it et al.}} Experimental one-way quantum computing, {\it Nature (London)} {\bf 434}, 169 (2005).
\bibitem{BBDRN09}Briegel, H. J., Browne, D. E., D\"{u}r, W., Raussendorf, R. \& Van den Nest, M. Measurement-based quantum computation, {\it Nat. Phys.} {\bf 5}, 19 (2009).
\bibitem{LJZHMDBBR13}Lanyon, B. P. {\red{\it et al.}} Measurement-Based Quantum Computation with Trapped Ions, {\it Phys. Rev. Lett.} {\bf 111}, 210501 (2013).
\bibitem{ABSRLRS18}A.-Arriagada, F. {\red{\it et al.}} One-way quantum computing in superconducting circuits, {\it Phys. Rev. A} {\bf 97}, 032320 (2018).
\bibitem{MF13}Morimae, T. \& Fujii, K. Blind quantum computation protocol in which Alice only makes measurements, {\it Phys. Rev. A} {\bf 87}, 050301(R) (2013).
\bibitem{BR01}Briegel, H. J. \& Raussendorf, R. Persistent Entanglement in Arrays of Interacting Particles, {\it Phys. Rev. Lett.} {\bf 86}, 910 (2001).
\bibitem{MDF17}Mantri, A., Demarie, T. F. \& Fitzsimons, J. F. Universality of quantum computation with cluster states and (X,Y)-plane measurements, {\it Sci. Rep.} {\bf 7}, 42861 (2017).
\bibitem{BFK09}Broadbent, A., Fitzsimons, J. \& Kashefi, E. Universal Blind Quantum Computation, in {\it Proceedings of the 50th Annual Symposium on Foundations of Computer Science} (IEEE, Los Alamitos, 2009), p. 517.
\bibitem{MP13}Mhalla, M. \& Perdrix, S. Graph States, Pivot Minor, and Universality of (X,Z)-measurements, {\it Int. J. Unconvent. Comput.} {\bf 9}, 153 (2013).
\bibitem{RHG06}Raussendorf, R., Harrington, J. \& Goyal, K. A fault-tolerant one-way quantum computer, {\it Ann. Phys.} {\bf 321}, 2242 (2006).
\bibitem{RHG07}Raussendorf, R., Harrington, J. \& Goyal, K. Topological fault-tolerance in cluster state quantum computation, {\it New J. Phys.} {\bf 9}, 199 (2007).
\bibitem{RH07}Raussendorf, R. \& Harrington, J. Fault-tolerant quantum computation with high threshold in two dimensions, {\it Phys. Rev. Lett.} {\bf 98}, 190504 (2007).
\bibitem{MF12}Morimae, T. \& Fujii, K. Blind topological measurement-based quantum computation, {\it Nat. Commun.} {\bf 3}, 1036 (2012).
\bibitem{AKLT88}Affleck, I., Kennedy, T., Lieb, E. H. \& Tasaki, H. Valence Bond Ground States in Isotropic Quantum Antiferromagnets, {\it Commun. Math. Phys.} {\bf 115}, 477 (1988).
\bibitem{BM08}Brennen, G. K. \& Miyake, A. Measurement-Based Quantum Computer in the Gapped Ground State of a Two-Body Hamiltonian, {\it Phys. Rev. Lett.} {\bf 101}, 010502 (2008).
\bibitem{MM16}Miller, J. \& Miyake, A. Hierarchy of universal entanglement in 2D measurement-based quantum computation, {\it npj Quantum Information} {\bf 2}, 16036 (2016).
\bibitem{GGM18}Gachechiladze, M., G\"{u}hne, O. \& Miyake, A. Changing the circuit-depth complexity of measurement-based quantum computation with hypergraph states, {\red {\it Phys. Rev. A} {\bf 99}, 052304 (2019)}.
\bibitem{KW17}Kissinger, A. \& van de Wetering, J. Universal MBQC with generalised parity-phase interactions and Pauli measurements, Preprint at arXiv:1704.06504 (2017).
\bibitem{HCDB07}Hartmann, L., Calsamiglia, J., D\"{u}r, W. \& Briegel, H. J. Weighted graph states and applications to spin chains, lattices and gases, {\it J. Phys. B} {\bf 40}, S1 (2007).
{\red\bibitem{HT19}Hayashi, M. \& Takeuchi, Y. Verifying commuting quantum computations via fidelity estimation of weighted graph states, Preprint at arXiv:1902.03369 (2019).}
\bibitem{HM15}Hayashi, M. \& Morimae, T. Verifiable Measurement-Only Blind Quantum Computing with Stabilizer Testing, {\it Phys. Rev. Lett.} {\bf 115}, 220502 (2015).
\bibitem{MK18}Markham, D. \& Krause, A. A simple protocol for certifying graph states and applications in quantum networks, Preprint at arXiv:1801.05057 (2018).
\bibitem{TMMMF18}Takeuchi, Y., Mantri, A., Morimae, T., Mizutani, A. \& Fitzsimons, J. F. Resource-efficient verification of quantum computing using Serfling's bound, Preprint at arXiv:1806.09138 (2018).
\bibitem{S02}Shi, Y. Both Toffoli and Controlled-NOT need little help to do universal quantum computation, Preprint at arXiv:quant-ph/0205115 (2002).
\bibitem{F86}Feynman, R. P. Quantum mechanical computers, {\it Found. Phys.} {\bf 16}, 507 (1986).
\bibitem{KSV02}Kitaev, A. Y., Shen, A. H. \& Vyalyi, M. N. {\it Classical and Quantum Computation} (AMS, Boston, 2002).
\bibitem{BL08}Biamonte, J. D. \& Love, P. J. Realizable Hamiltonians for universal adiabatic quantum computers, {\it Phys. Rev. A} {\bf 78}, 012352 (2008).
\bibitem{FK17}Fitzsimons, J. F. \& Kashefi, E. Unconditionally verifiable blind quantum computation, {\it Phys. Rev. A} {\bf 96}, 012303 (2017).
\bibitem{TFMI16}Takeuchi, Y., Fujii, K., Morimae, T. \& Imoto, N. Fault-tolerant verifiable blind quantum computing with logical state remote preparation, Preprint at arXiv:1607.01568 (2016).
{\red\bibitem{MK19}Morimae, T. \& Koshiba, T. Impossibility of perfectly-secure one-round delegated quantum computing for classical client, {\it Quantum Inf. Comput.} {\bf 19}, 214 (2019).
\bibitem{ACGK17}Aaronson, S., Cojocaru, A., Gheorghiu, A. \& Kashefi, E. On the implausibility of classical client blind quantum computing, Preprint at arXiv:1704.08482 (2017).
\bibitem{MNTT18}Morimae, T., Nishimura, H., Takeuchi, Y., \& Tani, S. Impossibility of blind quantum sampling for classical client, Preprint at arXiv:1812.03703 (2018).}
\bibitem{RHBM13}Rossi, M., Huber, M., Bru{\ss}, D. \& Macchiavello, C. Quantum hypergraph states, {\it New J. Phys.} {\bf 15}, 113022 (2013).
\bibitem{MTH17}Morimae, T., Takeuchi, Y. \& Hayashi, M. Verification of hypergraph states, {\it Phys. Rev. A} {\bf 96}, 062321 (2017).
\bibitem{TM17}Takeuchi, Y. \& Morimae, T. Verification of Many-Qubit States, {\it Phys. Rev. X} {\bf 8}, 021060 (2018).
\bibitem{ZH18}Zhu, H. \& Hayashi, M. Efficient verification of hypergraph states, Preprint at arXiv:1806.05565 (2018).
{\red\bibitem{M16}McKague, M. Interactive proofs for BQP via self-tested graph states, {\it Theory of Computing} {\bf 12}, 1 (2016).
\bibitem{MM10}McKague, M. \& Mosca, M. Generalized Self-testing and the Security of the 6-State Protocol, in {\it Proc. of Theory of Quantum Computation, Communication, and Cryptography 2010} (Springer, Leeds, 2011), p. 113.}
\end{thebibliography}
\end{document}